\documentstyle[12pt]{article}
\sf

\begin{document}
\input vatola.sty
\def\disp{\displaystyle}
\def\bm{\begin{minipage}[t]}
\def\em{\end{minipage}}
\def\srh{\stackrel{rightpoonup}}
\def\i{\int}
\def\s{\sum}
\def\f{\frac}
\def\p{\partial}
\def\bt{\begin{tabular}}
\def\et{\end{tabular}}
\def\bfr{\begin{flushright}}
\def\mm{\mbox{\boldmath $ }}
\def\efr{\end{flushright}}
\def\bfl{\begin{flushleft}}
\def\efl{\end{flushleft}}
\def\vs{\vspace}
\def\hs{\hspace}
\def\sta{\stackrel}
\def\pb{\parbox}
\def\bc{\begin{center}}
\def\ec{\end{center}}
\def\sp{\setlength{\parindent}{2\ccwd}}
\def\bp{\begin{picture}}
\def\ep{\end{picture}}
\def\uni{\unitlength=1mm}
\def\REF#1{\par\hangindent\parindent\indent\llap{#1\enspace}\ignorespaces}
%\columnsep 1.2 true cm

\noindent
\bc
{\LARGE\bf Analytic Solution of Ground State for\\
\vspace{.2cm}
Coulomb Plus Linear Potential}

\vs*{5mm}
{\large  W. Q. Chao(Zhao)$^{1,~2}$, C. S. Ju$^{2}$}

\vs*{11mm}

{\small \it 1. China Center of Advanced Science and Technology (CCAST)}

{\small \it (World Lab.), P.O. Box 8730, Beijing 100080, China}

{\small \it 2. Institute of High Energy Physics, Chinese Academy of Sciences,

P. O. Box 918(4), Beijing 100039, China}

\ec
%\newpage
\vspace{2cm}
\begin{abstract}

The newly developed single trajectory quadrature method is applied
to solve the ground state quantum wave function for Coulomb plus linear potential.
The general analytic expressions of the energy and wave function
for the ground state are given. The convergence of the solution is
also discussed. The method is applied to the ground state of
the heavy quarkonium system.
\end{abstract}
\vspace{.5cm}
\noindent
PACS{:~~11.10.Ef,~~03.65.Ge}

\newpage

Recently a new method has been developed by R. Friedberg, T. D. Lee
and W. Q. Zhao[1,2]
to solve the N-dimensional low-lying quantum wave functions of
Schr$\ddot{{\rm o}}$dinger
equation using quadratures along a single trajectory. Based on the
expansion on $1/g$, where $g $ is a scale factor expressing the strength
of the potential, Schr$\ddot{{\rm o}}$dinger equation can be cast into
a series of first order partial differential equations, which is further
reduced to a series of integrable first order ordinary differential
equations by single-trajectory quadratures. New perturbation series
expansion is also derived
based on this method, both for one-dimensional and N-dimensional cases.
Some examples for one-dimensional problems have been illustrated in
[1,2].

Coulomb plus linear potential has been widely applied to describe the heavy
quarkonium state. However, it is difficult to obtain an analytic expression of
the energy and wave function. Here the single trajectory quadrature method is
applied to solve the ground state for Coulomb plus linear potential. The general
analytic expressions of the energy and wave function
for the ground state are given. The convergence of the solution is
also discussed. The result
is applied to describe the ground state of heavy quarkonia. Some discussions of
the limitation of its applicability is given at the end.

Let us consider a unit mass particle moving in a central potential.
The Schr$\ddot{{\rm o}}$dinger equation in the 3-dimensional space is expressed as
\begin{eqnarray}\label{e1}
[-\f{1}{2}{\bf \nabla}^2+V(r)]\Psi (r)= E \Psi(r).
\end{eqnarray}
Based on the single trajectory quadrature method[2], following steps should be taken
to solve the problem for the ground state.

\noindent
1. For the potential $V(r)$, the scale factor $g$ is introduced as
\begin{eqnarray}\label{e2}
V(r)=g^k v(r).
\end{eqnarray}
First, the energy is expanded in terms of $1/g$:
\begin{eqnarray}\label{e3}
E = g^l E_0+g^{l-i}E_1+g^{l-2i}E_2+\cdots
\end{eqnarray}
The highest $g$-power, $l$, in the expansion could be fixed by the dimensional
consideration in the following way: Assume the behavior of the potential
$v(r)$ approaches $r^n$ when $r \rightarrow 0$ and $E_0$ is dimensionless.
The dimension of $\nabla^2$ in the first term of the right hand side in (\ref{e1}) is
the same as $[r^{-2}]$. The dimension of each term in (\ref{e1}), namely $[r^{-2}]$
for ${\bf \nabla}^2$, $[g^k r^n]$ for $V(r)$ and $[g^l]$ for $E$ should be the same.
This then gives
\begin{eqnarray}\label{e4}
l = \frac{2 k}{n+2}.
\end{eqnarray}

The ground state wave function $\Psi(r)$ is expressed as
\begin{eqnarray}\label{e5}
\Psi(r) = e^{- S(r)}.
\end{eqnarray}
Substituting (\ref{e5}) into (\ref{e1}) a equation for $S(r)$ and $E$ is obtained as
\begin{eqnarray}\label{e6}
\f{1}{2}{\bf \nabla}^2 S(r)-\f{1}{2}({\bf \nabla} S(r))^2 + V(r) - E = 0.
\end{eqnarray}
Then $ S(r)$ is also expanded in terms of $1/g$:
\begin{eqnarray}\label{e7}
S(r) = g^m S_0(r)+g^{m-j} S_1(r)+
g^{m-2j} S_2(r)+\dots
\end{eqnarray}
Substitute the expressions (\ref{e3}) and (\ref{e7}) into the equation (\ref{e6}).
By equating the
coefficients of each $g^{-n}$, a series of first order differential equations could
be obtained. Now the highest $g$-power, $m$, in the $S$-expansion should be
determined. The highest power of $g^{2m}$ comes from the second
term $-\f{1}{2}({\bf \nabla} S)^2 $ of the left hand side of
(\ref{e6}), which gives  $- \frac{1}{2}({\bf \nabla} S_0(r))^2$ to
the first one of the series of equations. The highest $g$-powers in
$V(r)$ and $E$ are $ g^k$ and $g^l$ respectively. When $k>l$,
$v(r)$ should enter the first equation, as in the case of Harmonic
Oscillator potential[1,2], which requires $2m=k$; in the second
equation for $g^{2m-j}$ power, ${\bf \nabla} S_0\cdot{\bf \nabla}S_1$
should be related to $E_0$, which gives $2m-j=l$. On the other
hand, when $k<l$, the first equation should be related to $E_0$
and this gives $2m=l$; the second equation with the power
$g^{2m-j}$ is then related to $v(r)$, which gives $2m-j=k$.
From the above discussion we derive the following condition:
\begin{eqnarray}\label{e8}
2m=k,&&~~2m-j=l,~~~~~{\sf for}~~l<k,\nonumber\\
2m=l,&&~~2m-j=k,~~~~~{\sf for}~~l>k.
\end{eqnarray}
To ensure that  $\{E_i\}$ and $\{ S_i\}$ enter the equations successively, we
have $i=j$ in (\ref{e3}) and (\ref{e7}).

For the Coulomb plus linear potential
\begin{eqnarray}\label{e9}
V(r) = g^2(- \frac{1}{r} + \mu r)
\end{eqnarray}
$k=2$ and
\begin{eqnarray}\label{e10}
v(r) = - \frac{1}{r} + \mu r,
\end{eqnarray}
which gives $n=-1$. From (\ref{e4}) and (\ref{e8}) this gives
$l=4>k$, which leads to $m=2$ and $i=j=2$. Therefore, for this potential (\ref{e3}) and
(\ref{e7}) are expressed as
\begin{eqnarray}\label{e11}
E &=& g^4 E_0 + g^2 E_1 + E_2 + g^{-2} E_3 + \cdots + g^{-(2n-4)} E_n
+ \cdots,\nonumber \\
S &=& g^2 S_0 + S_1 + g^{-2} S_2 + g^{-4} S_3 + \cdots + g^{-(2n-2)} S_n
+ \cdots.
\end{eqnarray}
Substituting (\ref{e11}) into equation (\ref{e6}), equating the coefficients of each
$g^n$ term, the following series of equations are obtained:
\begin{eqnarray}\label{e12}
({\bf \nabla} S_0)^2 &=& -2 E_0,\\
{\bf \nabla} S_0 \cdot {\bf \nabla} S_1 &=&
\f{1}{2}{\bf \nabla}^2 S_0 + v(r) - E_1,\\
{\bf \nabla} S_0 \cdot {\bf \nabla} S_2 &=&
\f{1}{2}{\bf \nabla}^2 S_1 -\f{1}{2}({\bf \nabla} S_1)^2
 - E_2,\\
&&\cdots\nonumber\\
{\bf \nabla} S_0 \cdot {\bf \nabla} S_n &=& \f{1}{2}{\bf \nabla}^2 S_{n-1}
- \frac{1}{2}\sum\limits_{m=1}^{n-1}{\bf \nabla} S_m \cdot {\bf \nabla}
S_{n-m} - E_n\\
&&\cdots\nonumber
\end{eqnarray}

\noindent
2. Following ref.[2] the series of equations (\ref{e12}-15) could be solved easily.
Considering ${\bf \nabla} S_0 = \f{d S_0}{d r}$ the equation (\ref{e12}) gives
\begin{eqnarray}\label{e16}
S_0(r) = \sqrt{-2 E_0}~r.
\end{eqnarray}
Substituting (\ref{e16}) into (13) for $E_1$ and ${\bf S}_1(r)$,
considering  ${\bf \nabla}^2 r=\frac{2}{r}$, we have
\begin{eqnarray}\label{e17}
\sqrt{-2 E_0} \frac{d S_1}{d r} =
\frac{1}{2} (\sqrt{-2 E_0}\frac{2}{r})+(-\frac{1}{r} + \mu~r) - E_1.
\end{eqnarray}

To keep $S_1(r)$ regular at $r =  0$ we have
\begin{eqnarray}\label{e18}
E_0 = -\frac{1}{2}~~~~~{\sf and}~~~~~~S_0(r) = r.
\end{eqnarray}
This gives
\begin{eqnarray}\label{e19}
{\bf \nabla}S_0 \cdot {\bf \nabla}S_1 = \frac{d S_1}{d r} &=& \mu r - E_1, \nonumber\\
S_1(r)&=& \frac{1}{2} \mu~r^2 - E_1~r.
\end{eqnarray}
Substituting (\ref{e19}) into (14) for $S_2(r)$ and $E_2$,
considering ${\bf \nabla}^2 r^2=6$, we have
\begin{eqnarray}\label{e20}
\frac{d S_2}{d r} = \frac{1}{2}(\frac{1}{2} \mu \cdot 6 - \frac{2E_1}{r}) -
\frac{1}{2}( \mu~r - E_1)^2 - E_2.
\end{eqnarray}
In order that  $S_2(r)$ be regular at $r=0$, we have
\begin{eqnarray}\label{e21}
E_1  = 0,~~~S_1(r) = \frac{1}{2}\mu~r^2.
\end{eqnarray}
This gives
\begin{eqnarray}\label{e22}
\frac{d S_2}{d r} &=& -\frac{1}{2}\mu^2 r^2 + \frac{3}{2} \mu
-E_2,\nonumber\\
S_2(r)&=&  -\frac{1}{6}\mu^2 r^3 + \frac{3}{2} \mu~r -E_2~r.
\end{eqnarray}
Following similar procedure for $S_3(r)$ and $E_3$,
introducing ${\bf \nabla}^2 r^3=12~r$, we have
\begin{eqnarray}\label{e23}
\frac{d S_3}{d r} = \frac{1}{2}( -2 \mu^2~r +
\frac{3\mu}{r}-\frac{2E_2}{r}) - \mu~r(-\frac{1}{2} \mu^2r^2 +\frac{3\mu}{2} -E_2)-E_3.
\end{eqnarray}
In order to have $S_3(r)$ also regular at
$r=0$, we get
\begin{eqnarray}\label{e24}
E_2  &=& \frac{3}{2} \mu\nonumber\\
S_2(r) &=& -\frac{1}{6} \mu^2 r^3.
\end{eqnarray}
Then (\ref{e23}) becomes
\begin{eqnarray}\label{e25}
\frac{d S_3}{d r} &=& - \mu^2~r + \frac{1}{2} \mu^2r^3-E_3,\nonumber\\
S_3(r) &=& - \frac{1}{2}\mu^2~r^2 + \frac{1}{8} \mu^2r^4-E_3~r.
\end{eqnarray}
Continuing similar argument we could reach
\begin{eqnarray}\label{e26}
E_3 = 0,~~~~~~~~~&& S_3(r) = -\frac{1}{2} \mu^2 + \frac{1}{8} \mu^3 r^4,\nonumber \\
E_4 = \frac{3}{2} \mu^2,&&S_4(r)= -\frac{3}{4} \mu^3 r^3 - \frac{3}{20} \mu^4 r^5,
 \\
&& \cdots\nonumber
\end{eqnarray}
Now we introduce the general expression
\begin{eqnarray}\label{e27}
\frac{d S_n}{d r} = \sum\limits_{0\leq l < \frac{n}{2}}
\alpha_l^{(n)} r^{n-2l} \mu^{n-l}
\end{eqnarray}
then we have
\begin{eqnarray}\label{e28}
S_n &=& \sum\limits_{0\leq l < \frac{n}{2}}\frac{1}{(n+1-2 l)}
\alpha_l^{(n)} r^{n+1-2l} \mu^{n-l}\nonumber\\
{\bf \nabla}^2 S_n = \frac{1}{r^2}~ \frac{d}{d r}r^2 \frac{d S_n}{d r}
&=& \sum\limits_{0\leq l < \frac{n}{2}}
\alpha_l^{(n)} (n+2-2 l) r^{n-1-2l} \mu^{n-l}~.
\end{eqnarray}
Substituting (\ref{e27}) and (\ref{e28}) into (15) and comparing
the coefficients of the same power of $r$ we obtain
\begin{eqnarray}\label{e29}
\alpha_l^{(n)} &=& \frac{1}{2} \alpha_{l-1}^{(n-1)}(n+3-2 l)\nonumber\\
&&-\frac{1}{2} \sum\limits_{m=1}^{n-1}
\sum\limits_{i \geq 0,~i>l-\frac{n-m}{2}}^{i\leq l,~i<\frac{m}{2}}
\alpha_i^{(m)}\alpha_{l-i}^{(n-m)}~.
\end{eqnarray}
To keep $S_{2n}$ or $S_{2n+1}$ regular at $r=0$ we have
\begin{eqnarray}\label{e30}
E_{2n-1}=0,~~~E_{2n}=\frac{3}{2}\alpha_{n-1}^{(2n-1)} \mu^n.
\end{eqnarray}
From $\alpha_0^{(1)}=1$ all the $\alpha_l^{(n)}$
and $E_n$ for $n>0$ could be derived based on (\ref{e29}) and (\ref{e30}).
Combining with $E_0=- \frac{1}{2}$ and $S_0=r$ we obtain
\begin{eqnarray}\label{e31}
E &=& g^4( -\frac{1}{2}+\frac{3}{2} \frac{\mu}{g^4} -\frac{3}{2} (\frac{\mu}{g^4})^2 + \cdots) \\
\Psi(r) &=& {\rm exp}[-g^2~r -\frac{1}{2} \mu~r^2 + \frac{1}{6g^2} \mu^2~r^3
+ \frac{1}{2g^4} \mu^2~r^2\nonumber\\
&& - \frac{1}{8g^4} \mu^3 r^4 - \frac{3}{4g^6} \mu^3~r^3 + \frac{1}{8g^6} \mu^4~r^5 + \cdots]~.
\end{eqnarray}
The same procedure can be performed if we define
\begin{eqnarray}\label{e33}
\epsilon = g^2 \mu
\end{eqnarray}
and solve the equation (\ref{e1}) for the potential
\begin{eqnarray}\label{e34}
V(r) = -\frac{g^2}{r} + \epsilon r.
\end{eqnarray}
The derivation has been given in ref.[2] and the result is exactly the same as (\ref{e31})
and (32), considering (\ref{e33}) we have:
\begin{eqnarray}\label{e35}
E &=& g^4(-\frac{1}{2}+\frac{3}{2}~ \frac{\epsilon}{g^6}
-\frac{3}{2}~(\frac{\epsilon}{g^6})^2 + \cdots) \\
\Psi(r) &=& {\rm exp}[-g^2~r -\frac{1}{2g^2} \epsilon~r^2 + \frac{1}{6g^6}
\epsilon^2~r^3+ \frac{1}{2g^8} \epsilon^2~r^2\nonumber\\
&&- \frac{1}{8g^{10}} \epsilon^3 r^4
- \frac{3}{4g^{12}} \epsilon^3~r^3 + \frac{1}{8g^{14}} \epsilon^4~r^5 + \cdots]~.
\end{eqnarray}
Introducing a parameter $\lambda=\epsilon/g^6=\mu/g^4$ the energy
could be expressed as
\begin{eqnarray}\label{e35b}
E=g^4(- \frac{1}{2}+\sum\limits_{n\geq 1} \frac{3}{2}\alpha_{n-1}^{(2n-1)} \lambda^n).
\end{eqnarray}
Introducing $e_n= \frac{3}{2}\alpha_{n-1}^{(2n-1)}$ the energy
could be expressed as
\begin{eqnarray}\label{e36b}
E=g^4(- \frac{1}{2}+\sum\limits_{n\geq 1} e_n \lambda^n).
\end{eqnarray}
This method can easily calculate the energy expansion series up to
any order of $n$. It gives the possibility to analyze the
convergence of the series in details.
The convergence of the expansion series of the energy $E$ depends on the
parameter $\lambda$. In fact, this series is an asymptotic one. For certain value of
$\lambda$ we could only reach a certain accuracy of the energy.
In Table I the ratio of $R_n=|e_n/e_{n-1}|$ for
different $n$ is listed. From (\ref{e35b}) we know that the ratio of the successive terms
in the energy expansion series is $\lambda |e_n/e_{n-1}|=\lambda R_n$.
It can be seen that the series would be
meaningful only when $\lambda< \frac{1}{R_n}$. This gives the
limitation of the applicability of this method. It also tells us
how accurate the final result could reach for a fixed value of
$\lambda$.
In table II for each order $n$, the $\lambda$ corresponding to $\lambda
R_n \sim 1$ is listed. The obtained energy $E/g^4$ for this special value
of $\lambda$ at each order of $n$ is also given, together with the
reached accuracy $e_n \lambda^n$.
For example, when $n=11$ the corresponding $\lambda=.052$ gives
$\lambda R_{11} \sim 1$.
It means that the correction term increases when $n>11$ and increases
further. This would finally give a divergent result. Up to $n=11$ the obtained
$e_n \lambda^n=0.00004$ which gives the accuracy of the obtained energy
$E/g^4 \cong -0.43443$ for $\lambda=0.052$.

Now consider a pair of heavy quark and antiquark with equal mass $m$ and color
charge $q$ and $-q$, moving in a Coulomb plus linear potential. In the
non-relativistic approximation the wave function $\psi(r)$ to describe the relative
motion of the quark pair satisfies the following Schr$\ddot{{\rm o}}$dinger
equation:
\begin{eqnarray}\label{e33a}
[-\f{1}{2\cdot m/2}\nabla^2-\frac{q^2}{r} +\kappa~r]\psi (r)= {\cal E} \psi(r),
\end{eqnarray}
where $\kappa$ is the strength of the linear potential. After a simple
transformation, (\ref{e33a}) becomes (\ref{e1}) and with
\begin{eqnarray}\label{e34a}
\f{1}{2} m q^2 &=& g^2,\nonumber \\
\f{1}{2} m {\cal E} &=& E,\nonumber \\
\f{1}{2g^2} m \kappa &=& \mu~~~{\sf or}~~~\f{1}{2} m \kappa =
\epsilon\nonumber\\
\lambda \equiv \f{4\kappa}{m^2 q^6}&=&\f{\epsilon}{g^6}=\f{\mu}{g^4}.
\end{eqnarray}
Substituting (\ref{e34a}) into (\ref{e35b}) the ground state energy
${\cal E}$ of the two quark system can be expressed as
\begin{eqnarray}\label{e35a}
{\cal E} &=& \frac{1}{2}m q^4(- \frac{1}{2}
+ \sum\limits_{n\geq 1} \frac{3}{2}\alpha_{n-1}^{(2n-1)}
(\f{4\kappa}{m^2 q^6})^n)\nonumber\\
&=& \frac{1}{2}m q^4(-\f{1}{2}+\f{3}{2} \lambda -\f{3}{2} \lambda^2
+ \frac{27}{4} \lambda^3\cdots).
\end{eqnarray}
For the quarkonium system the color charge $q^2$ can be related to the strong coupling
constant $\alpha_s$ as[3]
\begin{eqnarray}\label{e36}
q^2 = \frac{4}{3} \alpha_s.
\end{eqnarray}
For example, taking $q^2 = 0.5$ and $\kappa=1~GeV/fm$ we have
\begin{eqnarray}\label{e37}
\lambda=\frac{4\kappa}{m^2q^6} \sim 6.4 GeV^2 \times
\frac{1}{m^2}.
\end{eqnarray}
Taking the mass of the charm quark $m_c=1.6~GeV$, we have
\begin{eqnarray}\label{e38}
\lambda_c=\frac{4\kappa}{m_c^2q^6} \sim 2.5~,
\end{eqnarray}
which is too large for the application of this method.
If we look at the expansion series, for the ground state of $J/\psi$ we have
\begin{eqnarray}\label{e40}
m_{J/\psi} &=& 1.6~GeV \times 2 + {\cal E}_c\nonumber \\
           &\sim& 3.2~GeV + .2~GeV(-.5+3.7-9.4+105+\cdots).
\end{eqnarray}
Obviously, this expansion has a very bad behaviour. Therefore this method could not be
applied to charmnium since the charm quark mass is not heavy enough.
For bottom and top quark, taking $m_b \sim 4.5~GeV$ and  $m_t \sim 170~GeV$,
we could obtain
\begin{eqnarray}\label{e39}
\lambda_b=\frac{4\kappa}{m_b^2q^6} \sim 0.32~~~{\sf and}~~~
\lambda_t=\frac{4\kappa}{m_t^2q^6} \sim 2.2 \times 10^{-4}.
\end{eqnarray}
For the ground state of the bottom and top quarkonium we have
\begin{eqnarray}\label{e41}
m_{b\overline{b}} &=& 4.5~GeV \times 2 + {\cal E}_b\nonumber \\
           &\sim& 9.0~GeV + .56~GeV(-.5+0.48-0.15+0.22+\cdots)\\
m_{t\overline{t}} &=& 170~GeV \times 2 + {\cal E}_b\nonumber \\
           &\sim& 340~GeV + 21.3~GeV(-.5+3.3 \times 10^{-4}-7.3\times 10^{-8}+\cdots).
\end{eqnarray}
From (\ref{e41}) and (48) it can be seen that in the case of bottom quarkonium
$\lambda_b \sim 0.32$, the accuracy of the derived energy has been
improved and this method could well be applied to the top
quarkonium.\\

The authors would like to thank Professor T. D. Lee for his continuous
instructions and advice. This work is partly supported by NNSFC (No. 19947001).\\

\newpage

\bc
{\bf References}
\ec

1.  R. Friedberg, T. D. Lee, W. Q. Zhao, IL Nuovo Cimento A112(1999),1195

2.  R. Friedberg, T. D. Lee, W. Q. Zhao, Ann. phys. 288(2001),52

3.  Introduction to High-Energy Heavy-Ion Collisions, C. Y. Wong,

\noindent
~~~~~~World Scientific Pub., p344

\vspace{1cm}

\begin{center}
{\large Table I. $R_n=|e_n/e_{n-1}|$}
\end{center}

\begin{tabular}{|c|c|c|c|c|c|c|c|c|c|c|}
% after \\: \hline or \cline{col1-col2} \cline{col3-col4} ...
\hline
   n     & 3 & 4 & 5 & 6 & 7 & 8 & 9 & 10 & 11 & 12\\
\hline
$R_n$    & 4.5 & 7.36 & 9.67 & 11.62 & 13.35 & 14.95 & 16.45 & 17.90 & 19.32 & 20.72\\
\hline
$1/R_n$  & .22 & .14 & .10 & .086 & .075 & .067 & .061 & .056 & .052 & .048 \\
\hline
\end{tabular}

\begin{center}
{\large Table II. The obtained energy and its accuracy for different $\lambda$}
\end{center}

\begin{tabular}{|c|c|c|c|c|c|c|c|c|c|c|}
\hline
$n$ & 3 & 4 & 5 & 6 & 7 & 8 & 9 & 10 & 11 & 12 \\
\hline
$\lambda$ & .22 & .14 & .10 & .086 & .075 & .067 & .061 & .056 &
.052 & .048 \\
\hline
$E/g^4$ & .20 & .31 & .361 & .379 & .394 & .4048 & .4130 & .4198 &
.42535 & .43089\\
\hline
$|e_n| \lambda^n$ & .07 & .02 & .006 & .002 & .001 & .0004 & .0002 & .0001 &
.00004 & .00002\\
\hline
\end{tabular}

\end{document}